# *Ab initio* investigation of the effects of B-doping on the adsorption of $H_2O$, $H_2$ and $O_2$ molecules at diamond surfaces

Stefanos Giaremis[1], Maria Clelia Righi[1, *]

[1] Department of Physics and Astronomy, University of Bologna, 40127 Bologna, Italy

[*]**Corresponding Author**

Maria Clelia Righi – E-Mail: clelia.righi@unibo.it, Department of Physics and Astronomy, University of Bologna, 40127 Bologna, Italy.

## Abstract

Boron doped diamond is extensively studied for its use in tribological and electrochemical applications due to its remarkable physical and chemical properties. However, ambient conditions play a major role to its macroscopically observed behavior. In this study, the fundamental interactions between the low Miller index (001), (110) and (111) B-doped diamond surfaces with $H_2O$, $H_2$ and $O_2$ molecules, which are commonly present in ambient air and commonly involved in electrochemical reactions, are investigated by means of *ab initio* simulations. The results are presented in close comparison with previous studies on undoped diamond surfaces to reveal the impact of B on the adsorption properties. It is demonstrated that the B dopant is preferably incorporated on the topmost carbon layer and enhances the physisorption of $H_2O$ by forming a dative bond with O, while, in some cases, it can weaken the adsorption of $O_2$, compared to the undoped diamond. Moreover, a noticeable displacement of the surface atoms attached to the fragment of the dissociated $H_2O$ and $O_2$ molecules was observed, which can be associated to the first stage of wear at the atomistic level. These qualitative and quantitative results aim to provide useful insight towards the development of improved protective coatings and electrochemical devices.

## Introduction

Diamond is extensively used in a wide variety of applications in materials science for its remarkable properties. Specifically, due to its ultrahigh hardness and excellent overall mechanical properties, micro- and nanocrystalline diamond is commonly used in the form protective thin film coatings grown usually with techniques such as chemical vapor deposition (CVD), for increasing wear resistance, for instance against erosion [1], scratch [2] and corrosion [3]. Surface functionalization by nanostructuring can also lead to interesting antibacterial and antibiofouling properties [4]. An important application of diamond coatings, both in crystalline and in the amorphous form of diamond like carbon (DLC), is in tribological applications. Carbon-based coatings in general yield low values of coefficient of friction (COF) and wear rates [5]. However, the ambient conditions can deeply affect these tribological properties. In general, it is observed that the passivation of the carbon dangling bonds reduces the adhesive friction. Hydrogen termination was shown to reduce friction on diamond and DLC coatings, as well as dissociated $H_2O$ [6, 7, 8, 9, 10, 11]. For what concerns $O_2$ different results are reported in the literature: $O_2$ was found to increase friction and wear in some cases [12, 13, 14], but ultrathin nanodiamond films treated with oxygen plasma have been reported to present self-lubricating properties (useful also for their self-cleaning and optical properties) [15].

To further reduce COF and improve properties such as adhesion but also biocompatibility, the use of boron doped diamond is widely explored [16, 17, 18]. It has been demonstrated that B-doped nanocrystalline films exhibit lower coefficients of friction than undoped ones, with the presence of B also associated to increased $sp^3$-to-$sp^2$ ratio [19]. Moreover, as for the undoped diamond, lower friction and wear rates were observed under water lubrication, compared to dry ambient conditions in BDD films [20]. Appropriate surface texturing has been demonstrated to further improve the friction and wear behavior of BDD [17, 20]. It is also observed that the tribological properties of BDD are sensitive to the B doping level. In DLC films with boron concentrations of 2.9 at. %, the friction coefficient was reduced compared to undoped DLC, though at a slightly elevated wear rate [18]. A threefold increase in wear rate with increasing B content from 0.6 at. % to 2.8 at. %, along with a decrease of hardness and elastic modulus in micro and nanocrystalline diamond on Si substrates was reported experimentally [21]. The authors report the presence of amorphization on wear tracks after ball-on-surface sliding tests, which they propose that originates from weak lattice points due to the relatively weak B—B and B—C bonds. On the other hand, Lu *et al.* fabricated microcrystalline BDD on cobalt-cemented tungsten carbide (WC-Co) with hot filament CVD (HCVD), exploring a B content range up to 1.5 at. %, demonstrating that a B content of 0.8 at. % was associated to the lowest COF and highest hardness and wear resistance [22]. Moreover, they found that this doping level led to the orientation of grains to the (110) direction, and that B atoms introduced a suitable amount of tensile stress to compensate the compressive stress of the diamond coating.

Another major field of application for BDD is in electrocatalytic applications, due to its electronic properties. Undoped diamond, despite presenting some very attractive properties such as low dielectric constant, extreme thermal conductivity (2200 $Wm^{-1}K^{-1}$ at room temperature), electron and hole mobility (4500 $cm^2V^{-1}s^{-1}$ and 3800 $cm^2V^{-1}s^{-1}$, respectively, at room temperature) and dielectric breakdown field (5-10 $MVcm^{-1}$), has also a wide bandgap (5.47 eV) and low electrical conductivity, which hinders its potential use in electronic devices [23]. Doping with B leads to a p-type behavior, adding acceptor states at about 0.4 eV above the valence band [24], and this is the most used strategy for tuning its electronic properties, as B presents a very low carrier activation energy (0.37 eV [25]). Doping with B content below $10^{19} cm^{-3}$ leads to clear valence band. A semiconductor-to-metal transition was theoretically predicted to occur at a B content of $2\times10^{20}$ $cm^{-3}$ [26] and experimentally, fast electron transfer indicating metallic behavior was observed at $1 - 4.5 \times 10^{20}$ $cm^{-3}$ [27, 28, 29]. Moreover, since BDD presents other properties such as wide electrocatalytic window values (for instance, 4.9 V in ionic liquid, 4.6 V in organic solutions, 3.2 V in aqueous solutions) low, stable capacitance currents (3.6–7 μF cm−2 in aqueous, 14–20 μFcm$^{-2}$ in organic and 11–15 μFcm$^{-2}$ in ionic liquid solutions), high biocompatibility and chemical inertness in harsh environments (acidic or alkaline solutions, high voltages and current densities), it is very widely used in numerous applications in electrochemistry for electroanalysis, electrocatalysis/photoelectrocatalysis and electrosynthesis, biosensing and environmental purposes such as wastewater treatment and degradation of pollutants [30, 31, 32, 33, 34].

Atomistic scale effects are important for understanding the behavior of BDD in this type of applications. For instance, the presence of $sp^3$-bonded carbon was shown to be beneficial for the wide potential window due to the weak adsorption of intermediates during the $H_2$ and $O_2$ evolution reactions, and for the anticorrosion and antifouling properties due to the weak adsorption of other fouling species [35]. Moreover, the oxygen reduction reaction was proposed to occur preferably on the $sp^3$-bonded carbon component of electrodes [36]. High $sp^2$ concentration, on the other hand, was linked to a narrower window and surface oxidation [37]. Surface termination with H was linked to hydrophobicity, interfacial electron transfer, small potential window, and narrow bandgap with empty surface states while termination with O was linked to hydrophilicity and thus, higher background currents (due to the presence of water), wider potential windows and wide and clean bandgaps [34, 38, 39]. Hydrogen termination is known to lead also to the effect of negative electron affinity, i.e., the energy of the conduction band being higher than the vacuum level [40]. Hydroxyl is also used as surface termination, as a strategy for further reactions with other functional groups [41].

From the above, it is evident that the macroscopic properties of diamond and DLC coatings arise from the complex interplay of several factors, such as ambient conditions, crystallinity and grain size, bonding properties and the interaction strength between interfaces or surfaces and molecules. Therefore, assessing the impact of such atomistic-level mechanisms is important to the determination of the macroscopic behavior of BDD in fields such as the above. For this reason, atomistic simulations have been extensively used to unveil qualitatively and quantitatively relevant underlying phenomena. For instance, *ab initio* calculations have been used to compare the adsorption of water and hydrogen on the (2×1) Pandey reconstructed and unreconstructed diamond (111) surfaces, quantifying the adsorption energetics and demonstrating that the latter becomes energetically preferable to the former for surface coverage greater than about 40% due to chemisorbed adsorbents [42]. Similarly, the coverage of the (001) surface from the dissociative adsorption of atmospheric molecules and the associated potential energy surfaces were studied in [43]. The adsorption of water, hydrogen, and oxygen on the hydrogen and oxygen-terminated unreconstructed and Pandey reconstructed (111) surface orientations was investigated in [44], while the adsorption of the aforementioned molecules on the Pandey (001), (110) and (111) surfaces was examined in [45]. Density functional theory (DFT) calculations were also used to propose a theoretical model for the oxidation etching of the diamond (001) surface via the adsorption of the $O_2$ [46]. Recently, *ab initio* molecular dynamics simulations were used for examining the tribochemical reactions of $H_2O$, $H_2$ and $O_2$ on the diamond (111) and silica interface [47], as well as the water lubrication of sliding diamond surfaces through hydroxylation due to tribological effects [48]. The impact of B doping to the adsorption energies of hydrogen, oxygen, hydroxyl, and fluorine as surface-terminating species was explored by Zhao *et al*. [49], while the adsorption of the aforementioned molecules on the clean, B-doped (001) surface was investigated in [50]. The adsorption of water and hydroxyl radicals on hydrogen-terminated (111) surfaces was considered in [51]. *Ab initio* molecular dynamics on the interaction of the O- and H- terminated (111) and (001) BDD with water molecules was also used to assess the main factors of reactivity in each case [52].

However, while the adsorption of $H_2O$, $H_2$ and $O_2$ is extensively investigated, as shown above, for various undoped and B-doped surface orientation with various coverage strategies, a fundamental analysis of the adsorption of these molecules on the clean, B-doped (001), (110) and (111) surfaces is not yet reported in the literature. Therefore, in this work, the results of DFT calculations are presented for the molecular and dissociative adsorption cases mentioned above, in close comparison to previous theoretical results for the respective undoped surfaces [45]. Such information has been shown to be invaluable for extracting conclusions for further macroscopic properties, such as wettability [53], which can subsequently motivate new strategies in areas such as thin films coating development for tribological applications, or improve, for instance, the current understanding of reaction pathways in electrochemical applications.

## Computational Details

All the DFT calculations were performed by using the QUANTUM ESPRESSO package [54] with the Vanderbilt ultrasoft pseudopotentials [55] based on the Perdew – Burke – Ernzerhof derivation of the generalized gradient approximation for the expression of the exchange correlation functional [56] and a cut-off energy of 30 Ry for the expansion of the plane-wave basis set. The semi-empirical D2 dispersion correction by Grimme [57, 58] was applied to account for the long-range interactions between the surface and the considered molecules in the physisorption states. This level of theory has been previously used in similar adsorption studies for undoped (e.g. [45, 42, 46]) and B-doped diamond surfaces (e.g. [50, 51]) as a suitable trade-off between accuracy and computational cost. Moreover, calculations with the PBE functional were demonstrated to yield the same trends in the adsorption energies of the $H_2O$, $H_2$ and $O_2$ molecules and a similar level of accuracy compared to the PBEsol and PBE0 funtionals, for the adsorption of the aforementioned molecules in the undoped C(001), C(011) and Pandey C(111) surfaces [45], and also better agreement with experimental results compared to local-density approximation and Perdew-Wang 91 functionals [50]. Due to the presence of unpaired electrons from surface dangling bonds and molecular fragments, spin polarization was accounted for in all cases. A Gaussian smearing with a width of 0.01 Ry was applied for the integration of the Brillouin zone, which was sampled by a 2×2×1 Monkhorst-Pack generated k-point mesh [59]. The convergence thresholds for the total energy and the ionic forces were set to $10^{-4}$ Ry and $10^{-3}$ Ry/Bohr, respectively, while the convergence threshold for the electronic self-consistent loop was set to $10^{-5}$ Ry. The most stable configurations for each adsorption case were determined by sampling a set of both of high and low symmetry positions for the initial positions of the corresponding molecule or fragments, respectively, and then comparing the total energies of the respective relaxed configurations. Consequently, adsorption energies (per molecule), $E_{ads}$, for both molecular and dissociative adsorptions, were calculated via $E_{ads} = E_{tot} - E_{surf} - E_{mol}$, where $E_{tot}$ is the total energy of the adsorbate system, $E_{surf}$ is the total energy of the clean, relaxed surface and $E_{mol}$ is the energy of an isolated $H_2O$, $H_2$ or $O_2$ molecule, calculated on a 20×20×20 cell. Visualizations were produced with XCrySDen [60].

Four different diamond surface configurations were considered, namely the C(001), the C(110), the unreconstructed C(111) and the Pandey reconstructed C(111), represented in orthorhombic periodic cells with a minimum lateral dimension of 8.75 Å, a minimum thickness of 6 atomic layers and a minimum vacuum size of 15 Å along the direction normal to the surface, to ensure that interactions between periodic images along this direction are negligible. The (2×1) dimer reconstruction was used for the C(001) surface, while for the case of the (111) surface, both the unreconstructed and the Pandey reconstructed cases were considered, since the reconstructed case is most likely to occur when the surface is not passivated (or, for instance, passivated with a hydrogen coverage below approximately 40% [42]), but it was also computationally demonstrated that B as a dopant on the surface can have a stabilizing effect on the unreconstructed (111) surface [61]. The choice of cell dimensions was similar to [45] for the C(001), C(110) and Pandey C(111) configurations and similar to [47] for the unreconstructed C(111) configuration, and the simulation parameters were demonstrated to yield an energy error smaller than 3 meV. To investigate the local effects of B doping on atomistic level, on each configuration, one C atom was substituted with a B one, leading to a B concentration of 0.5-1.0 at. %. This choice is based on previous evidence that B in general favors substitutional doping [62], while more specifically, it has also been demonstrated from electron energy loss spectroscopy measurements that in the above concentration levels, B is found in the form of single and isolated substitutional atoms [63]. The respective concentration range used here is commonly referred to as "normal" or "medium" doping concentration level [64], and as mentioned above, is often used in studies regarding both tribological and electrochemical behavior.

# Results and Discussion
## Structural properties of B-doped diamond surfaces
To investigate the energetically preferable site of the B dopant, substitutional sites at the three topmost (with respect to the exposed surface) atomic layers of each surface orientation were considered. The resulting configurations, along with the energy difference of each case compared to the energetically preferable one for each orientation, are presented in Figure 1. For the (001), (110) and Pandey reconstructed (111) surfaces, B atom is preferable positioned on the topmost atomic layer. For these cases, sites deeper within the crystal lattice can be considered as metastable and will likely lead to the consequent diffusion of the dopant.

In the (001) surface, the B dopant on its most stable site is positioned on the surface dimer. Since B has only 3 electrons (compared to the 4 electrons of C), it forms single C—B bonds compared to the double C=C ones in the rest of the dimer chains. This is also evident from the breakdown of the C=C double bond (1.38 Å) to single B—C (1.57 Å) at the topmost layer. By evaluating the corresponding Löwdin partial charges, the B atom was found to be positively charged by +0.27e on its preferable position, while its three neighboring C atoms are negatively charged by -0.09e. This effect is reasonable considering the lower electronegativity of B compared to C (2.04 versus 2.55 Pauling units, respectively [65]). These results agree with previous *ab initio* studies that also confirmed the tendency of B to occupy $sp^2$-bonded sites at the topmost surface layer [50, 66].

In the (110) surface, it can be assumed from their respective bond length (1.44 Å), that dangling bonds at the surface C atoms are being saturated by the formation of π-like or hybrid C—C chains. As in the previous case, the B dopant is also most stable at the topmost atomic layer for the (110) surface, with the same distribution of Löwdin partial charges between the dopant and its surrounding C atoms as in the (100) surface (+0.27e for B and -0.09e for its surrounding C atoms). The B—C bond length in the most stable site is 1.51 Å, so in this case, the preference of B atom for the topmost site can be also explained in terms of stress accommodation, as the elongation of the bonds that it is associated to, in this case, matches more closely the bond length of the diamond structure.

For the case of the Pandey reconstructed (111) surface, B atom prefers the site on the dimer chain, following the same trend as in the (001) surface. The B—C bond length in the dimer chain is 1.53 Å, while the respective one for the C—C bond in the chain is 1.43 Å. The Löwdin partial charge attributed to B in this case is +0.15e. The two surrounding C atoms in the chain dimer present a partial charge of -0.09e. This slight imbalance in partial charges, besides the known limitations of this scheme, can be also attributed to a small amount of charge transfer from the trench C atoms in the second atomic layer towards the ones in the surface chain dimers, as for the undoped surface, the former yields a value of +0.04e and the later a value of +0.01e. The presence of B on the topmost dimer chain was also associated with the lowest value of surface energy in a previous *ab initio* study [61]. The bond lengths in the top C—C dimer chains for the (001), (110) and Pandey reconstructed (111) surface are in agreement with previous *ab initio* works at the same level of theory [45].

On the other hand, in the case of the unreconstructed (111) surface, the B atom is most stable at the third atomic layer, as also previously shown from *ab initio* simulations [51]. For this surface orientation, in all the B-doped cases, as well as in the undoped one, the top bilayer appears slightly compressed, with a C—C bond length of 1.50 Å, while its bonds with the third atomic layer are slightly elongated at 1.65 Å. This slight tendency of the unreconstructed (111) surface towards graphitization is also reported in a previous *ab initio* study [51]. The tendency of the B atom to be substitutionally placed towards bulk lattice sites was also previously computationally demonstrated [66]. In its less stable sites at the first two atomic layers, B atom is found to form B—C bonds with a length of 1.53 and 1.52 Å, respectively, while in the most stable site at the third atomic layer, it actually forms bonds that match very closely the bond length of diamond (1.55 Å), while the respective C—C bond length in the same atomic layer in the undoped surface was 1.54 Å. In the latter case, the C atom on the second atomic layer that lies on top of the B dopant, is found to be shifted by 0.36 Å compared to the rest of the C atoms in the same layer, leading to a separation distance of

2.10 Å from the B atom underneath it. This effect can be explained by the fact that B has 3 electrons, therefore it forms three covalent bonds with the three neighboring C atoms in the fourth atomic layer, but it is weakly bonded with its fourth neighboring C atom on second atomic layer on top of it. The former, consequently, forms shorter, π-like bonds, with a length of 1.45 Å, with its three neighboring C atoms on the surface layer. This local bonding effect was also observed in [51]. The associated Löwdin partial charge of the B atom was found to be +0.16e at its most stable site at the third layer and +0.32e at the topmost site.

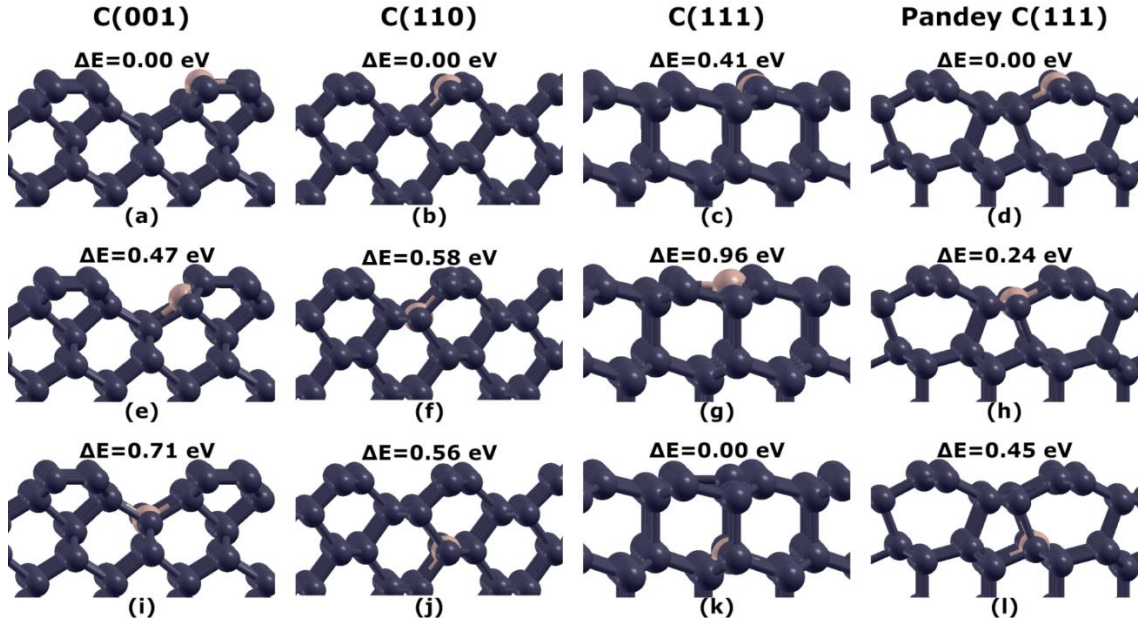

Figure 1: Relaxed configurations of the B-doped surface models with the B atom on the first (a-d), second (e-h) and third (i-l) atomic layers from the surface, along with their respective energy difference, ΔE, compared to the energetically preferable configuration for each surface orientation, for the four considered surface orientations. Pink spheres correspond to the B dopants while grey spheres correspond to C atoms.

## Molecular adsorption on B-doped diamond surfaces

The most stable configurations for the molecular adsorption of the $H_2O$, $H_2$ and $O_2$ molecules on the B-doped diamond (001), (110) unreconstructed and Pandey reconstructed (111) surfaces, with the B dopant on the topmost layer are presented in Figure 2. However, as mentioned above, for the case of the unreconstructed (111) surface, the B dopant is preferably positioned in the third atomic layer. Moreover, as the case of the undoped unreconstructed C(111) surface was not considered in [45], the cases of the adsorption of the $H_2O$, $H_2$ and $O_2$ molecules on the unreconstructed C(111) surface with the B dopant on the third atomic layer (denoted as "B3") and the undoped unreconstructed C(111) surface, where explicitly investigated for completeness, and the respective results of molecular adsorption are presented in Figure 3. The adsorption energies of the above cases are demonstrated in comparison with the ones of the respective undoped surfaces from [45] in Figure 6a, while the average adsorption energies for each molecule over all the considered surface orientations are presented in Table IV.

## Molecular adsorption of $H_2O$

The most evident impact of the presence of the B dopant on the topmost layer of all the considered diamond surfaces is towards the molecular adsorption of the $H_2O$ molecule, as in this case, the surface-molecule interaction takes the form of chemisorption, while for the respective undoped surfaces, it has been computationally demonstrated to lead to a weak physisorption [45, 42]. As previously computationally demonstrated but only for the (001) surface [50], the surface B atom tends to create a dative bond with the O atom of the $H_2O$ molecule, which is associated to a pronounced energy gain due to the filling of the B outer shell. From Figure 2 (a-d), it is evident that this also holds for the cases of the (110) and unreconstructed (111) orientations. The corresponding value of the adsorption energy per molecule, $E_{ads}$, for the (001) surface was found to be more slightly more pronounced, compared to the respective value for the (110) and the unreconstructed (111) configurations, while also the bond length between B and O follows the same trend (Table I). From the Löwdin partial charge analysis (Table I), it is evident that in all cases the B atom becomes less positively charged after the interaction with the $H_2O$ molecule, while the O atom becomes less negatively charged.

Moreover, for the case of the unreconstructed (111) surface, when B is in the third atomic layer (which was found to be its most stable site), the molecular adsorption of $H_2O$ becomes effectively similar to the respective case for the undoped surface. The adsorption energy per molecule for the former case (-0.20 eV) can even potentially indicate a slightly less favorable adsorption tendency compared to the latter one (-0.23 eV). The results for the undoped case are also in good agreement with a previous ab initio study [42]. The distance between the lowermost H atom of the $H_2O$ molecule and the topmost atom of the BDD surface was 2.35 Å in the former and 2.37 Å in the latter case. The adsorption energy for the undoped case is also in good agreement with the results of Levita *et al.* (-0.22 eV) [42], while the weakening of the adsorption interaction between BDD and $H_2O$ with increasing depth of the B dopant in this surface configuration was also previously reported from *ab initio* simulations with the PBE-D2 approach [51].

The molecular adsorption of water on the Pandey reconstructed (111) surface is weaker compared to the previous cases, with the interaction being closer to a physisorption. This is evident by the lower energy gain, ($E_{ads}$ = -0.43 eV for the B-doped versus -0.23 eV for the undoped surface), and the longer distance between the B and O atoms in the most stable configuration (1.77 Å). The Löwdin partial charge of the B atom is virtually the same before and after the interaction with the $H_2O$ molecule (Table I).

Overall, from the adsorption energies of the above cases, it is evident that the molecular adsorption of $H_2O$ is stronger on the (001) surface of BDD, while for the undoped diamond, it has been demonstrated that the adsorption was stronger on the (110) surface [45]. The order of reactivity for this type of adsorption on the B-doped surfaces was (001) > (110) > unreconstructed (111) (B1) > Pandey (111) > unreconstructed (111) (B3). Furthermore, by considering the average adsorption energies on all surface orientations (Table IV), it is evident that the impact of B is mostly pronounced in this type of adsorption, leading to an average increase of the respective energy gain by almost 0.7 eV with respect to the undoped case. The increased energy gain for the molecular adsorption of $H_2O$ can be also associated to an increased probability of a $H_2O$ molecule being captured from the surface. The attraction of $H_2O$ molecules towards the B-doped diamond surface may be favor the surface oxidation with negative effects on its resistance to wear. On the contrary the increased surface hydrophilicity can be linked to the lower coefficients of friction observed under water lubrication [20], and to the attractive properties of BDD for use in electrochemical applications such as wastewater treatment [32].

Table I: Values of Löwdin partial charges of B and O before and after adsorption, and B—O bond length for the most stable configurations of molecular adsorption of $H_2O$ on the B-doped (001), (110), unreconstructed and Pandey reconstructed (111) surfaces.

|  | (001) | (110) | (111)-B1 | Pandey (111) |
|---|---|---|---|---|
| **B partial charge (before adsorption) (e)** | +0.27 | +0.27 | +0.32 | +0.15 |
| — ||— — ||— (after adsorption) (e) | +0.24 | +0.16 | +0.20 | +0.16 |
| **O partial charge (before adsorption) (e)** | -0.73 | -0.73 | -0.73 | -0.73 |
| — ||— — ||— (after adsorption) (e) | -0.56 | -0.44 | -0.45 | -0.49 |
| **B—O (Å)** | 1.61 | 1.62 | 1.65 | 1.77 |

## Molecular adsorption of $H_2$

As for the case of the $H_2O$ molecule, the molecular adsorption of $H_2$ from the undoped diamond surface has been previously computationally demonstrated to take the form of physical adsorption [45]. The incorporation of B atom on the topmost layer changes this picture for the cases of the (001) and the unreconstructed (111) surfaces (Figure 3 e-h). In these cases, the adsorption becomes stronger, with the associated adsorption energies per molecule appearing increased, compared to the ones for the corresponding adsorptions on the undoped surfaces (-0.44 eV versus -0.07 eV from [45] for the (001) surface and -0.17 eV versus -0.07 eV from [45] for the unreconstructed (111) surface). For the (001) surface, there is a pronounced attraction of the $H_2$ molecule from the B dopant, as the configuration in its most stable state was found to lead to a chemisorption state by the formation of a dative B—H bond and the stretching of the H—H bond (Table II). The adsorption appears less pronounced for the case of the unreconstructed (111) surface with the B dopant on the surface layer, as the energy gain was found to be less increased and a longer B—H dative bond was formed (Table II). The stretching of the H—H bond was also less significant in this case (Table II). In both cases, from the partial charge analysis, it was found that B was became less positively charged while H became more positively charged after adsorption, indicating a potential charge transfer towards B (Table II).

Table II: Values of Löwdin partial charges of B and H before and after adsorption, and B—H and H—H bond lengths for the most stable configurations of molecular adsorption of $H_2$ on the B-doped (001) and unreconstructed (111) surfaces.

|  | (001) | (111)-B1 |
|---|---|---|
| **B partial charge (before adsorption) (e)** | +0.27 | +0.32 |
| — ||— — ||— (after adsorption) (e) | -0.04 | +0.10 |
| **H partial charge (before adsorption) (e)** | +0.04 | +0.04 |
| — ||— — ||— (after adsorption) (e) | +0.26 | +0.20 |
| **B—H (Å)** | 1.40 | 1.57 |
| **H—H (Å)** | 0.84 | 0.78 |

The $H_2$ molecule was found to be physisorbed on the other two surfaces. The adsorption energy for the BDD (110) surface was slightly weaker compared to the undoped case, while the molecule was found to adsorb at 2.78 Å from the surface (measuring from the lowest H atom to the topmost surface atom). Finally, in the Pandey reconstructed (111) surface, the adsorption energy on the doped surface was the same as in the undoped one, leading also to a weak physisorption, with the $H_2$ molecule in both cases being on top of the middle of a trench formation at 2.42 Å from the topmost surface atom.

Overall, the molecular adsorption of $H_2$ is weaker compared to the one of $H_2O$, which was also found to be the case for the respective undoped diamond surfaces [45]. However, the adsorption of $H_2$ was stronger on the (001) surface, as was also the case with $H_2O$, while for the undoped diamond surface the adsorption of $H_2$ was stronger on the (110) surface [45]. By also considering the average adsorption energies in Table IV, this type of adsorption is the weakest among the considered cases, with the addition of B leading to a small increase of energy gain by only approximately 0.08 eV in average.

**Molecular adsorption of $O_2$**

In contrast with $H_2O$ and $H_2$, the molecular adsorption of $O_2$ from undoped diamond surfaces has been demonstrated to have a mostly chemical character [45]. Here, the presence of B as dopant in the topmost atomic layer is also found to lead to the chemisorption of the $O_2$ molecule in all the considered cases, with the O═O double bond being rearranged into single bonded O—O, C—O and B—O configurations.

Specifically, in the (001) surface, the $O_2$ molecule is preferably positioned on top of the B—C heterodimer, in similar manner to the case of the undoped surface [45]. The molecular adsorption of oxygen was found to be weaker compared to the undoped surface [45] by 1.24 eV. The adsorption of $O_2$ on the B-doped (110) surface was also found to occur in a similar manner to the undoped case [45], with $O_2$ being preferably chemisorbed in the middle of the trench and each O atom attached to either B or a C atom at the opposite side of the trench. The adsorption on the B-doped (110) surface was found to be weaker compared to the respective undoped one [45] by 0.9 eV.

The impact of surface B-doping on the $O_2$ adsorption was less pronounced for the two cases considered for the (111) orientation, as the calculated adsorption energies were found to be very close to their respective counterparts for the undoped configurations. For the Pandey reconstructed surface, $O_2$ was found to be favorably positioned on top of B and its neighboring C atom on the surface chain. In the unreconstructed (111) configuration, when B was positioned on its preferable site in the third atomic layer, $O_2$ was preferably chemisorbed far from the vicinity of B. As for the assessment of the local impact of B, only sites up to the second coordination shell of the dopant were considered, the preferable position of $O_2$ for the molecular adsorption was found to be in a manner such as one O atom was on top of a C atom in the second coordination shell of B and the second O further away from B (fourth coordination shell). This configuration leads to a weaker adsorption of the molecule compared to the undoped case. However, by moving the $O_2$ molecule further away from the vicinity of B, the adsorption energy tends to reach the value of the undoped configuration.

The results of the partial charge analysis in Table III indicate charge transfer towards $O_2$, which however occurs through the C atom in which $O_2$ is attached to, as the change of the partial charge of B before and after adsorption is less pronounced. Moreover, the B—O length was found to be longer in all cases compared the respective C—O one (Table III), as expected.

Overall, the molecular adsorption of $O_2$ was found to be stronger on all the considered B-doped surfaces, compared to the molecular adsorption of $H_2O$ and $H_2$, as was also the case for the respective undoped configurations [12]. Furthermore, again as in the undoped case, the (110) configuration was found to be more favorable towards this type of adsorption. However, the second most favorable surface was the unreconstructed (111) (B1) for the B-doped case but the (001) for the undoped one. The adsorption on the Pandey reconstructed (111) surface was the least favorable both in the B-doped and the undoped configuration, as was also the case for the rest of the considered molecules, in agreement with previous studies demonstrating the inert character of this surface [42]. The molecular adsorption of $O_2$ was found to be the strongest among the considered types of molecular adsorption, with the addition of B leading to a reduction of the average energy gain by 0.55 eV per molecule (Table IV). However, the large values of standard deviation reflect the impact of the different surface orientations mentioned above.

Table III: Values of Löwdin partial charges of B, O attached to B ($O_B$), C and O attached to C ($O_C$), before and after adsorption, and B—O, C—O and O—O bond lengths for the most stable configurations of molecular adsorption of $O_2$ on the B-doped (001) and unreconstructed (111) surfaces.

|  | (001) | (110) | (111)-B1 | Pandey (111) |
|---|---|---|---|---|
| **B partial charge (before adsorption) (e)** | +0.27 | +0.27 | +0.32 | +0.15 |
| — ||— — ||— (after adsorption) (e) | +0.22 | +0.20 | +0.20 | +0.19 |
| **$O_B$ partial charge (before adsorption) (e)** | +0.05 | +0.05 | +0.05 | +0.05 |
| — ||— — ||— (after adsorption) (e) | -0.18 | -0.20 | -0.21 | -0.22 |
| **C partial charge (before adsorption) (e)** | -0.08 | +0.01 | +0.04 | -0.09 |
| — ||— — ||— (after adsorption) (e) | +0.09 | +0.23 | +0.25 | +0.10 |
| **$O_C$ partial charge (before adsorption) (e)** | +0.05 | +0.05 | +0.05 | +0.05 |
| — ||— — ||— (after adsorption) (e) | -0.12 | -0.15 | -0.26 | -0.14 |
| **B—O (Å)** | 1.51 | 1.47 | 1.49 | 1.55 |
| **C—O (Å)** | 1.47 | 1.46 | 1.44 | 1.49 |
| **O—O (Å)** | 1.48 | 1.49 | 1.59 | 1.48 |

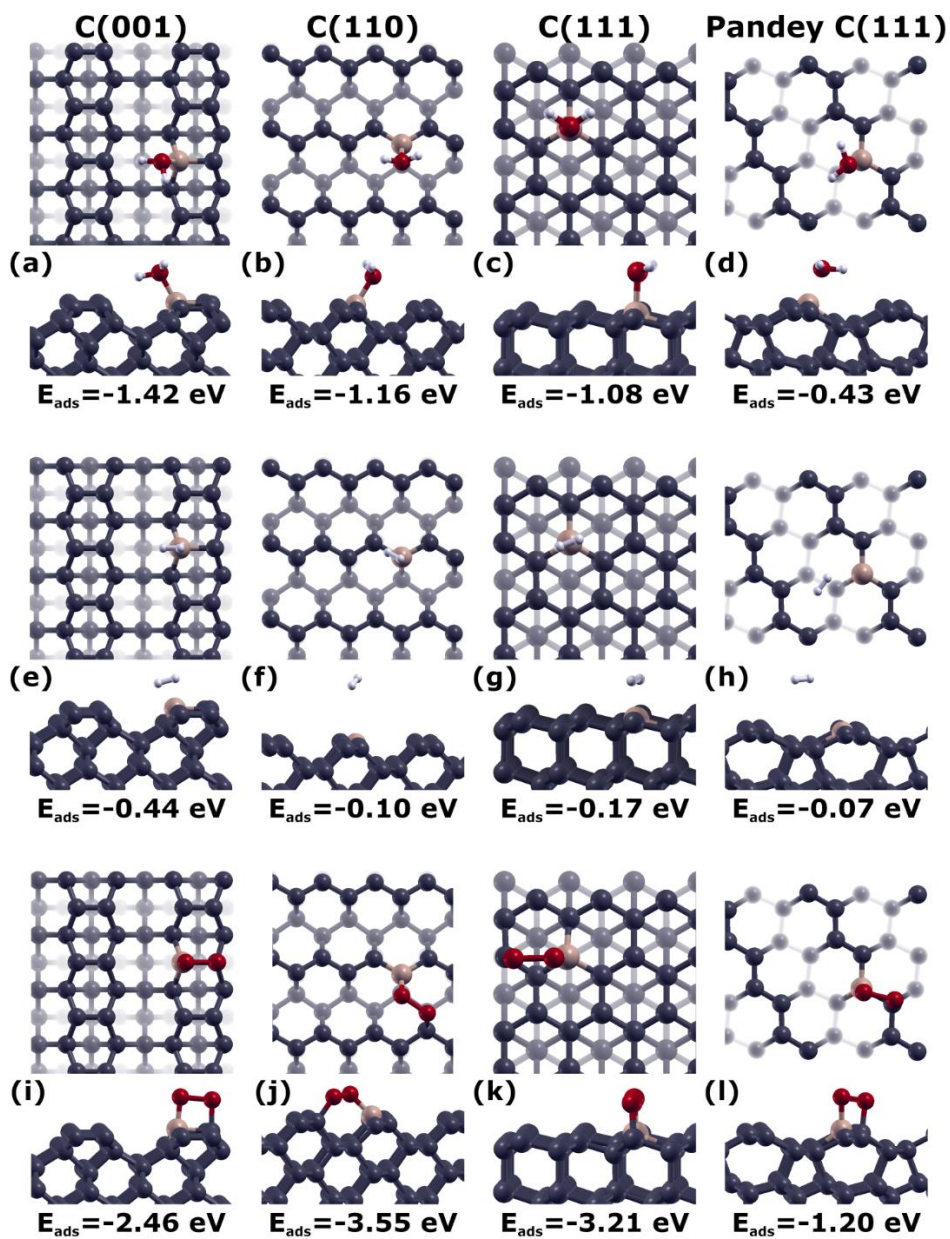

**Figure 2:** Top and side view representations of the most stable configurations for the molecular adsorption of the $H_2O$ (a-d), $H_2$ (e-h) and $O_2$ (i-l) molecules, along with their respective adsorption energies per molecule, for the four considered surface orientations. Pink spheres correspond to the B dopants while grey spheres correspond to C atoms. H and O atoms are represented with white and red spheres, respectively. In all cases, the B dopant is positioned on the topmost atomic layer.

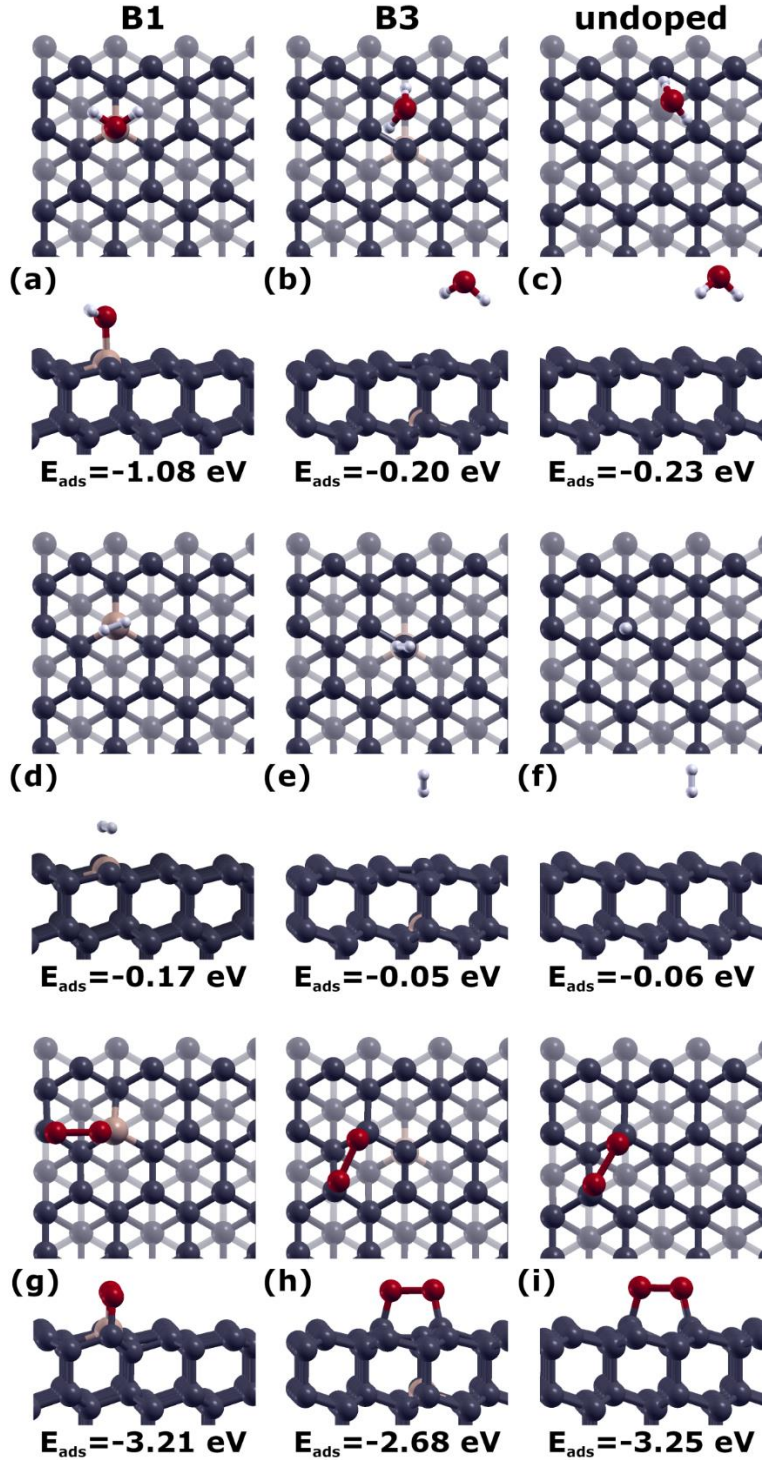

Figure 3: Top and side view representations of the most stable configurations for the molecular adsorption of the $H_2O$ (a-c), $H_2$ (d-f) and $O_2$ (g-i) molecules, along with their respective adsorption energies per molecule, for the B-doped, unreconstructed C(111) surface with the B atom on the first ("B1") and third ("B3") atomic layer, and the undoped unreconstructed C(111) surface. Pink spheres correspond to the B dopants while grey spheres correspond to C atoms. H and O atoms are represented with white and red spheres, respectively.

## Dissociative adsorption on B-doped diamond surfaces

The most stable configurations for the fragments of $H_2O$, $H_2$ and $O_2$ dissociated on the B-doped surfaces having B in the first layer are presented in Figure 4, along with the corresponding adsorption energies. For the unreconstructed (111) surface both the cases with B on the first ("B1") and ("third") atomic layers are considered and compared with the undoped one in Figure 5. The dissociative adsorption energies are compared to those obtained for the undoped surfaces [45] in Figure 6b. Overall, the impact of boron doping was less pronounced for the dissociative adsorption than for molecular adsorption.

## Dissociative adsorption of $H_2O$

As in the undoped cases, the hydrogen and hydroxyl fragments occurring as products from the dissociative adsorption were found to be preferably adsorbed from the topmost chains and dimers, with the hydroxyl group being preferably adsorbed on the B atom by forming a dative bond, and the H atom being adsorbed by its adjacent C atom. This was observed in all configurations except for the unreconstructed (111) surface. In the latter, when B atom was positioned on the topmost layer ("B1" configuration), both fragments were found to be preferably attached to C atoms in the first coordination shell of B, while when B atom was positioned in its most stable position in the third layer ("B3" configuration), the hydroxyl fragment was preferably attached further away from the site on top of the B dopant. The value of the adsorption energy for the B1 case was found to be close to the one for the undoped one, which was also the lowest compared to the rest of the considered surfaces. The results for the undoped surface are also in agreement with previous *ab initio* calculations [42]. The adsorption energy for the most stable configuration for the B3 case was higher, indicating a weaker adsorption on the surface near B, and thus the detrimental impact of B towards the $H_2O$ dissociative adsorption in this case. However, this effect was local, as placing the fragments further away from the site on top of the B dopant, the value of the adsorption energy was found to reach rapidly the corresponding value for the undoped case.

In all the configurations, the OH group was preferably oriented pointing the H away from the H fragment due to the electrostatic repulsion between them. The strength of the adsorption was found to be relatively similar for the (001), (110) and the unreconstructed (111) orientations, with the energy gain decreasing for unreconstructed (111) (B1) > (110) > (001) > unreconstructed (111) (B3) > Pandey reconstructed (111). The B—O bond lengths reflect the above trend, with values of 1.40 – 1.43 Å for the unreconstructed (111), (110) and (001) surfaces and 1.48 Å for the Pandey reconstructed (111) surface. The adsorption on the Pandey reconstructed surface was weaker by almost 2 eV, demonstrating again the inert character of this configuration. The addition of B appears to favor the dissociative adsorption of $H_2O$ in the Pandey configuration, as the energy gain was increased from 0.63 eV for the undoped surface to 1.18 eV for the B-doped one. The incorporation of B did not lead to any significant changes to the adsorption energies in the rest of the surface configurations, compared to their undoped counterparts. However, from Figure 4(a-d), it is observed that the adsorption of the hydroxyl group on the B atom in the (001), (110) and Pandey reconstructed (111) surfaces leads to a displacement of the latter. This effect was less pronounced in similar calculations on the undoped surfaces [45], and it can be potentially associated to evidence of increased wear in BDD (e.g., [19]).

Overall, the dissociative adsorption of $H_2O$ was found to be the weakest, compared to the dissociative adsorption of $H_2$ and $O_2$, both for the B-doped and the undoped surface orientations. The weak interaction of BDD with OH is also reported experimentally and is assumed in reaction pathways models for water oxidation, where hydroxyl intermediates are assumed to react weakly with the surface of the BDD electrodes [67, 68].

**Dissociative adsorption of $H_2$**

In the dissociative adsorption of $H_2$, the H fragments adsorb on the surface on the topmost chains and atomic layers in all the considered surfaces, forming single bonds with a length of 1.10 Å. In all cases, the H fragments were not preferably attached to the B dopant, but to its adjacent or next-to-adjacent C atoms. Specifically, the H atoms were preferably attached to the neighboring C atoms of the B dopant in the (110) and Pandey reconstructed (111) surfaces. In the (001) surface, one H atom was preferably adsorbed by the C atom on the B—C dimer and the other was adsorbed by a C atom on the adjacent dimer, rearranging the C═C double bond into a single C—C one. Lastly, in the unreconstructed (111) surface, H atoms were attached to C atoms in the second coordination layer of the B dopant.

Overall, the addition of B to the topmost layer was not found to have a very significant impact to this type of adsorption, as it was associated with a small decrease of the adsorption energy gain for the (001), (110) surface and unreconstructed (111) surfaces and an increase of the energy gain for the Pandey reconstructed (111) surface. The order of the adsorption energy gain for the dissociative adsorption of $H_2$ was found to be unreconstructed (111) (B1) > (110) > (001) > unreconstructed (111) (B3) > Pandey reconstructed (111).

The adsorption was found to be stronger on the unreconstructed (111) surface, with an energy gain of 4.12 eV for the B1 configuration. The very high energy gain for the unreconstructed (111) towards the $H_2$ dissociative adsorption agrees with the well-known ability of passivating this surface with H [69]. This effect is here demonstrated to hold also for the B-doped unreconstructed (111) surface. However, as in the case of the dissociative adsorption of $H_2O$, the incorporation of B in the third atomic layer in the unreconstructed (111) surface was found to lead to a reduction of the energy gain, when $H_2$ is adsorbed on the vicinity of B.

Moreover, even though $O_2$ was found to be most preferably adsorbed in all the considered B-doped and undoped surfaces compared to $H_2$ and $H_2O$, the dissociation of $H_2$ on the unreconstructed (111) surface was found to lead to the highest energy gain, compared to the dissociation of the other two considered molecules (Figure 6b). The fact that $H_2$ is shown to be preferably dissociated on the unreconstructed (111) surface compared to $O_2$ might be interesting for applications in tribology or catalysis, in which the adsorption of oxygen species would be undesired.

**Dissociative adsorption of $O_2$**

The dissociation of $O_2$ from the B-doped surfaces was found to occur in different ways, depending on the considered surface orientation, with the formation of ether-group or carbonyl-like configurations. The effect of B doping is evident in this type of adsorption, as in the undoped surfaces, the dissociation of $O_2$ was found to lead only to the formation of C—O—C ether configurations [45].

In particular, the formation of carbonyl configurations was observed in the B-doped (001) and the unreconstructed (111) surfaces. In the former case, the O fragments were found to break the B—C dimer and form double bonds with B and C. The O═B bond was observed to be longer (1.27 Å) than the O═C one (1.21 Å). The adsorption energy in this case was very close to the one of the respective undoped surface, even though in the former case, the adsorption was most stable in an ether configuration [45]. In the latter case, the adsorbed O fragments formed double-like O═C bonds with the dangling bonds of the C atoms in the second coordination shell of B, for the case of B on the surface (B1), and in the second and fourth coordination shells in the case of B in the third layer (B3), following the same pattern as in both previous cases of dissociative adsorption. The adsorption energy on the B1 case was again very close to the undoped one, but a higher value was calculated for B3. The O═C bond length in all the unreconstructed (111) surfaces was measured close to 1.35 Å.

On the other hand, the formation of ether configurations was found to be preferable for the (110) and Pandey reconstructed (111) surfaces. In both cases, the most stable configurations were similar to the respective undoped ones from [45]. Specifically, in the (110) surface, the O fragments were attached to C atoms, in bridge sites connecting two topmost chains, while in the Pandey (111) surface, both O atoms where adsorbed in bridge positions between two consecutive B—C and C—C pairs on the same chain. The adsorption in the B-doped cases was weaker than in the undoped one for these two surfaces by approximately 2 eV.

Overall, as was also demonstrated for the undoped diamond [45], the dissociative adsorption of $O_2$ is the strongest among the considered molecules and adsorption types, with the adsorption energy decreasing with (001) > (110) > unreconstructed (111) > Pandey (111), both for the undoped and the B-doped configurations. Moreover, from Figure 4(i-l), it is evident that the adsorption of the O fragments leads to the displacement of the surface atoms attached to them, mostly in the (001) and (110) configurations. Similar to the case of the dissociative adsorption of $H_2O$, this observation, along with the high energy gain of the $O_2$ adsorption (which is in general observed both for the B-doped and the undoped diamond surfaces), could also be associated to previous evidence of higher wear rates in BDD, compared to undoped diamond coatings [19]. Moreover, the formation of C—O—C pairs was experimentally found to be preferably formed in longer oxidation times compared to C—OH, along with evidence of surface reconstruction in the former case [41], in agreement with the results of adsorption energies and most stable geometries presented here.

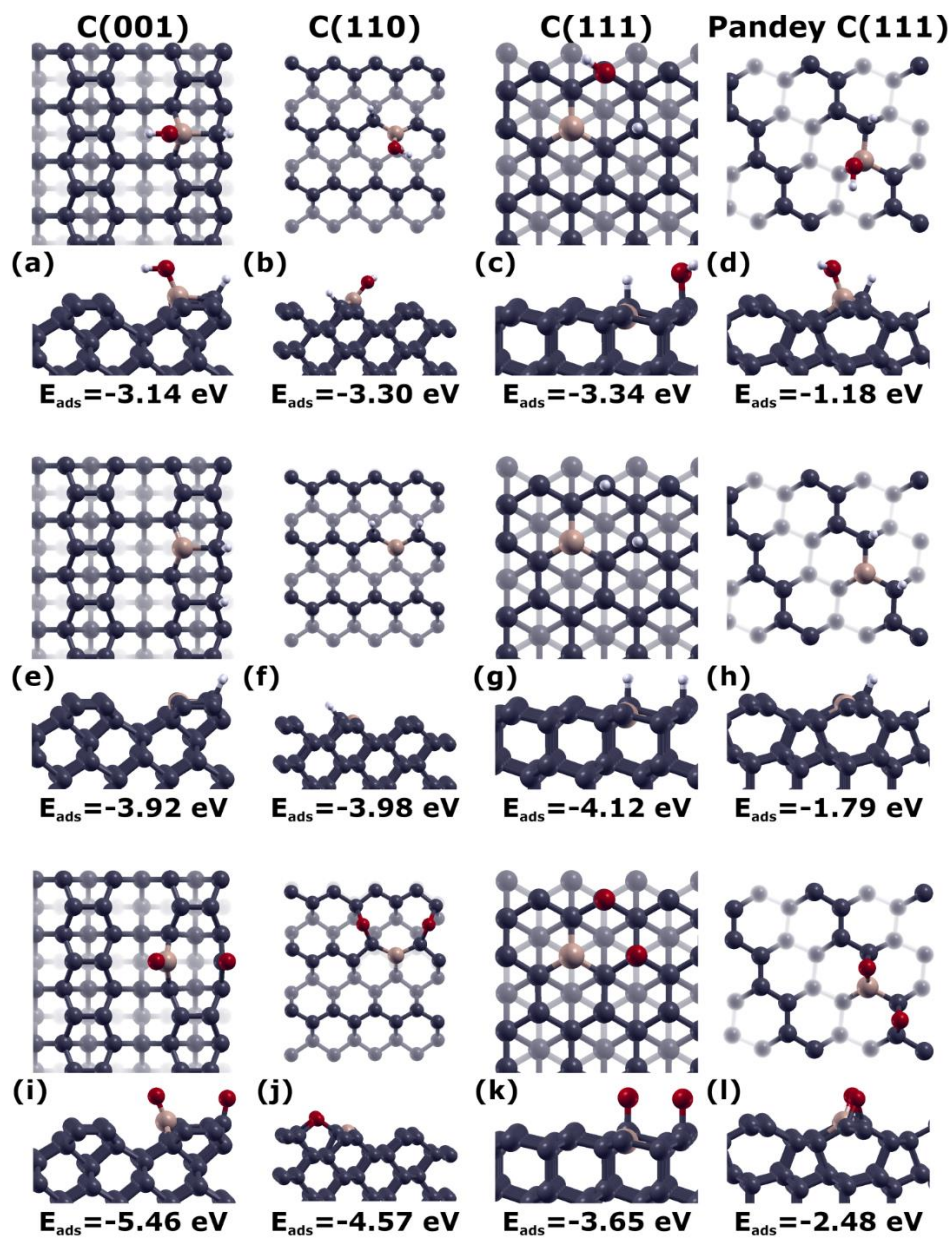

**Figure 4:** Top and side view representations of the most stable configurations for the dissociative adsorption of the $H_2O$ (a-d), $H_2$ (e-h) and $O_2$ (i-l) molecules, along with their respective adsorption energies per molecule, for the four considered surface orientations. Pink spheres correspond to the B dopants while grey spheres correspond to C atoms. H and O atoms are represented with white and red spheres, respectively. In all cases, the B dopant is positioned on the topmost atomic layer.

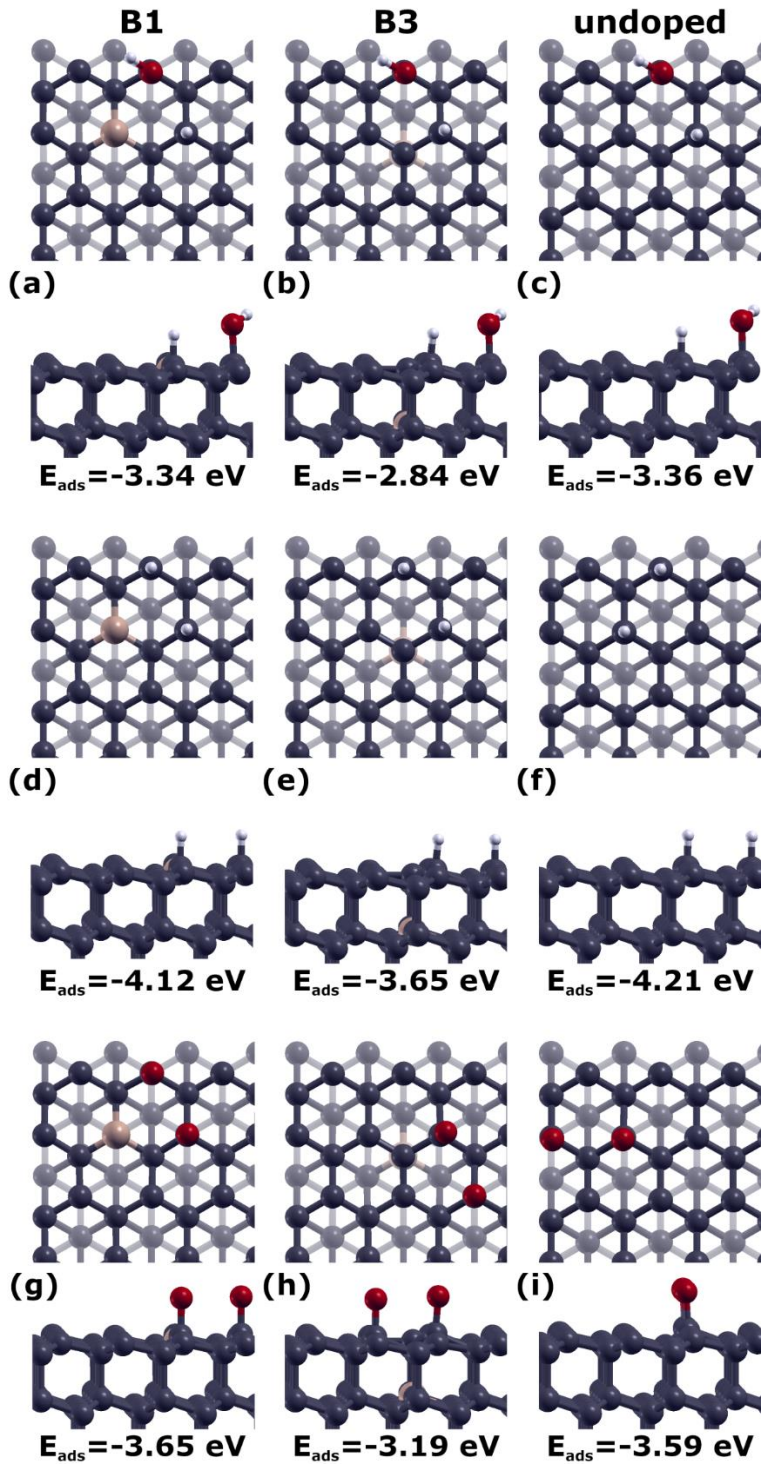

**Figure 5:** Top and side view representations of the most stable configurations for the dissociative adsorption of the $H_2O$ (a-c), $H_2$ (d-f) and $O_2$ (g-i) molecules, along with their respective adsorption energies per molecule, for the B-doped, unreconstructed C(111) surface with the B atom on the first ("B1") and third ("B3") atomic layer, and the undoped unreconstructed C(111) surface. Pink spheres correspond to the B dopants while grey spheres correspond to C atoms. H and O atoms are represented with white and red spheres, respectively.

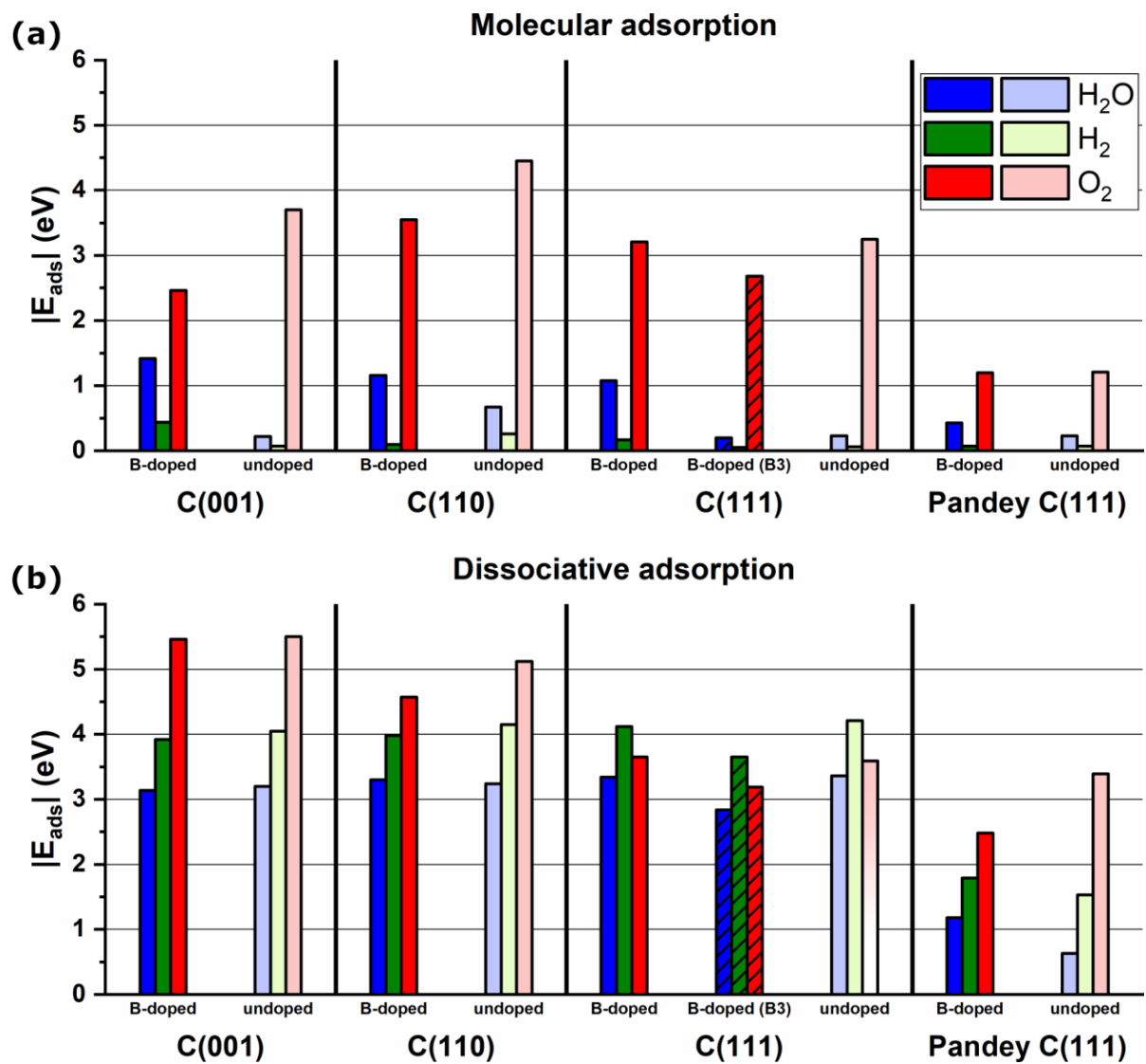

Figure 6: Absolute values of the adsorption energies per molecule of the most stable configurations for molecular (a) and dissociative (b) adsorption of the $H_2O$, $H_2$ and $O_2$ molecules on the four considered surface orientations. Cases denoted as "B-doped" correspond to surfaces with the B atom on the topmost atomic layer, except for the case of C(111) surface, where also the case of the B atom in the third atomic layer ("B3") is included. Results for $H_2O$ are represented with blue, for $H_2$ with green and for $O_2$ with red color. White gradient color coding corresponds to the cases of the undoped surfaces, while stripe-patterned bars in the case of the C(111) surface correspond to the "B3" case. Results for the undoped C(001), C(110) and Pandey reconstructed C(111) surfaces are taken from [45].

Table IV: Average adsorption energies per molecule, along with the respective values of standard deviation in parentheses (all in eV), for each molecule, estimated over all the four considered surfaces. Results are presented both for the molecular and the dissociative adsorption of the considered molecules.

|  | Molecular adsorption | | Dissociative adsorption | |
| --- | --- | --- | --- | --- |
|  | B-doped | undoped | B-doped | undoped |
| $H_2O$ | 1.02 (0.36) | 0.34 (0.19) | 2.75 (0.91) | 2.61 (1.14) |
| $H_2$ | 0.20 (0.15) | 0.12 (0.08) | 3.45 (0.96) | 3.49 (1.13) |
| $O_2$ | 2.61 (0.90) | 3.15 (1.20) | 4.04 (1.10) | 4.40 (0.92) |

# Conclusions

In this work, the molecular and dissociative adsorption of $H_2O$, $H_2$ and $O_2$ on boron doped (001), (110), and (111), both Pandey-reconstructed and unreconstructed, surfaces were studied *ab initio*, in close comparison with previous results for the corresponding undoped configurations [45]. In all cases, the B dopant was found to be preferably located in a substitutional site of the topmost layer, except for the unreconstructed (111), in which it was preferably located on the third atomic layer (in agreement with previous calculations [51]). In the latter case, it was found to reduce the adsorption strength in the vicinity of its topmost atom.

The presence of B on the considered diamond surfaces was shown to lead to the formation of a dative bond with O in the molecular adsorption of $H_2O$, increasing the energy gain for adsorption by 0.7 eV on average, and consequently increasing the probability for $H_2O$ molecules to be captured on the surface with respect to the undoped case. This result was in agreement with a previous *ab initio study* on the B-doped (001) surface [50], and can be associated to the previously observed improved friction coefficient of BDD under water lubrication [20, 22]. On the other hand, the molecular adsorption of $O_2$ was found to be weaker on the B-doped (001) and (110) surfaces compared to the undoped ones, while on the Pandey (111), less reactive, is similar in both the cases.

Moreover, the dissociative adsorption of $O_2$ was also found to be slightly weaker on the (110) and Pandey (111) B-doped surfaces. In these configurations, O atoms were found to be most stable in ether-like, bridge positions, while on the (001) and unreconstructed (111) surfaces, O atoms were found at carbonyl-like, on-top positions and the adsorption strength was similar to the undoped surfaces. In the most stable adsorption configurations for O dissociation on the (001) and (110) surfaces, a pronounced displacement of the surface atoms attached to the O adsorbents was observed, which could potentially act as the source of atomistic wear mechanisms. In the dissociative adsorption of $H_2O$, the hydroxyl fragments were preferably attached to the B atom and the H fragments to a C atom in the first coordination shell of B in most cases, and a pronounced displacement of surface atoms attached to the adsorbed fragments was also observed in the (001), (110) and Pandey (111) surfaces. The Pandey (111) was again found to be the less reactive surface towards the dissociation of all the considered molecules.

These results aim to provide a quantitative and qualitative understanding of the impact of boron doping on the atomistic mechanisms of adsorption of $H_2O$, $H_2$ and $O_2$ on the most common, diamond surfaces. The present theoretical analysis of the fundamental mechanisms of adsorption will hopefully improve the controlled use of B-doping in tribological, electrochemical and other technological applications of diamond.

## Acknowledgements

These results are part of the "Advancing Solid Interface and Lubricants by First Principles Material Design (SLIDE)" project that has received funding from the European Research Council (ERC) under the European Union's Horizon 2020 research and innovation program (Grant agreement No. 865633). Furthermore, we acknowledge the CINECA award under the ISCRA initiative, for the availability of high-performance computing resources and support.